\documentclass[3p,times,twocolumn]{elsarticle}

\usepackage{ecrc}

\usepackage{amsmath} 
\usepackage[none]{hyphenat} 

\volume{00}

\firstpage{1}

\journalname{Nuclear and Particle Physics Proceedings}

\runauth{}

\jid{nppp}

\jnltitlelogo{Nuclear and Particle Physics Proceedings}

\usepackage{amssymb}

\usepackage[figuresright]{rotating}

\begin{document}

\begin{frontmatter}



\dochead{}

   \title{Production of muons from
   heavy-flavour hadron decays at forward rapidity in Pb--Pb collisions
   at $\sqrt{s_{\rm NN}} = 5.02~{\rm TeV}$}


\author{Zuman~Zhang $^{1,2}$ for the ALICE Collaboration}
\address{$^1$ Key Laboratory of Quark and Lepton Physics (MOE) and Institute of Particle Physics, CCNU, Wuhan, China\\
$^2$ Laboratoire de Physique Corpusculaire, Clermont-Ferrand, France\\}

\begin{abstract}
  The measurement of the production of
  single muons from heavy-flavour hadron decays at forward rapidity in Pb--Pb
  collisions at $\sqrt{s_{\rm NN}} = 5.02~{\rm TeV}$ collected in 2015 is presented as a function of transverse momentum ($p_{\rm T}$) and collision centrality.
  A strong suppression of the yield is observed at high $p_{\rm T}$ in the most central collisions compared to
  the binary-scaled expectation from pp collisions at the same energy.
  The $p_{\rm T}$-integrated nuclear modification  factor ($R_{\rm AA}$) as function of the number of participating nucleons
  indicates an increase of the suppression from peripheral to central collisions. Comparisons with the results for Pb--Pb collisions at $\sqrt{s_{\rm NN}} = 2.76~{\rm TeV}$ and with transport model predictions are shown.
  A similar suppression is measured at both $\sqrt{s_{\rm NN}} = 2.76~{\rm TeV}$ and 5.02 TeV.
  The results available from $p_{\rm T} =$7 GeV/$c$ up to $p_{\rm T} =~$20 GeV/$c$, will provide new constraints on
  transport model ingredients and new insights on the understanding of
  the evolution of the hot and dense matter formed in ultra-relativistic heavy-ion
  collisions.
\end{abstract}


\end{frontmatter}



\section{Introduction}
\label{}
  ALICE~\cite{pub:ALICE} is the dedicated heavy-ion experiment at the LHC, optimized to investigate the
  properties of strongly-interacting matter under extreme conditions of temperature and energy
  density where the formation of the Quark-Gluon Plasma (QGP) is expected. Heavy quarks (charm
  and beauty) are regarded as efficient probes of the properties of the QGP as they are created on
  a very short time scale in initial hard parton scattering processes and subsequently interact with
  the medium. Open heavy flavours are expected to be sensitive to the energy density
  of the system through the in-medium energy loss of heavy quarks. In the high $p_{\rm T}$ region, the
  suppression of the heavy-favour yields is quantified by means of the nuclear modification factor,
  $R_{\rm AA}$:


  \begin{equation}
   R_{\rm AA}(p_{\rm T}) \equiv \frac{{\rm d}^{2}N_{\rm AA}/{\rm d}p_{\rm T}{\rm d}y}{<T_{\rm AA}>\,{\rm d}^{2}\sigma_{\rm pp}/{\rm d}p_{\rm T}{\rm d}y}
  \end{equation}

  It is defined as the ratio of the particle yield measured in Pb--Pb collisions to the
  $p_{\rm T}$-differential cross section measured in pp
  collisions scaled by the average nuclear overlap function.
  Due to the color-charge effect, the radiative energy
  loss of gluons should be larger than that of quarks. Moreover, 
  heavy-quark energy loss may be reduced with respect to that of light quarks due to the dead-cone effect~\cite{pub:deadcone}.

  \begin{flushleft}
    \bf{ALICE detector and data sample}
  \end{flushleft}

  We focus on the measurement of the nuclear modification factor of muons from open
  heavy-flavour hadron decays with the ALICE muon spectrometer in the pseudo-rapidity range of $-4 < \eta < -2.5$.
  The ALICE muon spectrometer consists of a thick front absorber, a beam shield, a dipole magnet, five tracking stations and two trigger stations behind an iron wall.
  In addition to the muon spectrometer, for the analysis presented here, the V0 detector comprising two scintillator arrays and the Silicon Pixel Detector (SPD) are used. Collisions were classified according to their centrality, determined from the sum of the amplitudes of the signals in the V0 detector and defined in terms of percentiles of the total hadronic Pb--Pb cross section. The event vertex is reconstructed with the SPD.

  The analysis uses the Pb--Pb data sample collected in 2015 at $\sqrt{s_{\rm NN}} = 5.02~{\rm TeV}$ and is based on muon triggered events.
  The trigger condition consists of a signal in the two V0 arrays (minimum bias trigger) with
  at least one candidate track with a high $p_{\rm T}$ threshold of about $4.2~{\rm GeV}/c$ in the muon trigger system.
  The integrated luminosity is 202 $\pm ~6 ~\mu \rm b^{-1} $ in the 0--90\% centrality class.

  Muons are reconstructed within the acceptance of the muon spectrometer. Then, it is required that the track candidate in the tracking system matches a track reconstructed in the muon trigger
  system. This condition allows us to reject most of the punch-through hadrons that are stopped
  in the iron wall. Furthermore, the correlation between
  the track momentum and the geometrical distance of closest approach to the primary vertex
  is used in order to remove fake tracks and tracks from beam-gas interactions.
  The remaining background for $p_{\rm T} >2 ~{\rm GeV}/c$ after these selection cuts consists of muons from decays of primary light hadrons ($\pi$ and K, main contribution at low and intermediate $p_{\rm T}$, called decay muons in the following) and muons from decays of W and Z/$\gamma^{*}$ (main contribution at high $p_{\rm T}$).

\section{Analysis strategy}

  \begin{equation}\label{eq:c4efficiencyCorrection}
   \frac{{\rm d}N_{\rm AA}^{\mu\leftarrow HF}}{{\rm d}p_{T}}=
   \frac{{\rm d}N_{\rm AA}^{incl~\mu}}{{\rm d}p_{T}}-
   \frac{{\rm d}N_{\rm AA}^{\mu\leftarrow K,\pi,W,Z/\gamma^{*} }}{{\rm d}p_{T}}
   \end{equation}


   The analysis strategy in Pb-Pb collisions includes the following steps:
   normalization of muon samples to equivalent number of minimum-bias (MB) events on a run by run basis, acceptance $\times$ efficiency correction and background subtraction (Eq. 2). 
   The acceptance $\times$ efficiency correction is obtained from realistic GEANT3 simulations using $p_{\rm T}$ and rapidity distributions of muons from heavy-flavour signals by NLO pQCD calculations. The centrality dependence of the
   tracking efficiency is estimated via an embedding procedure.

  \begin{figure}
    \centering
    \includegraphics[width=8cm]{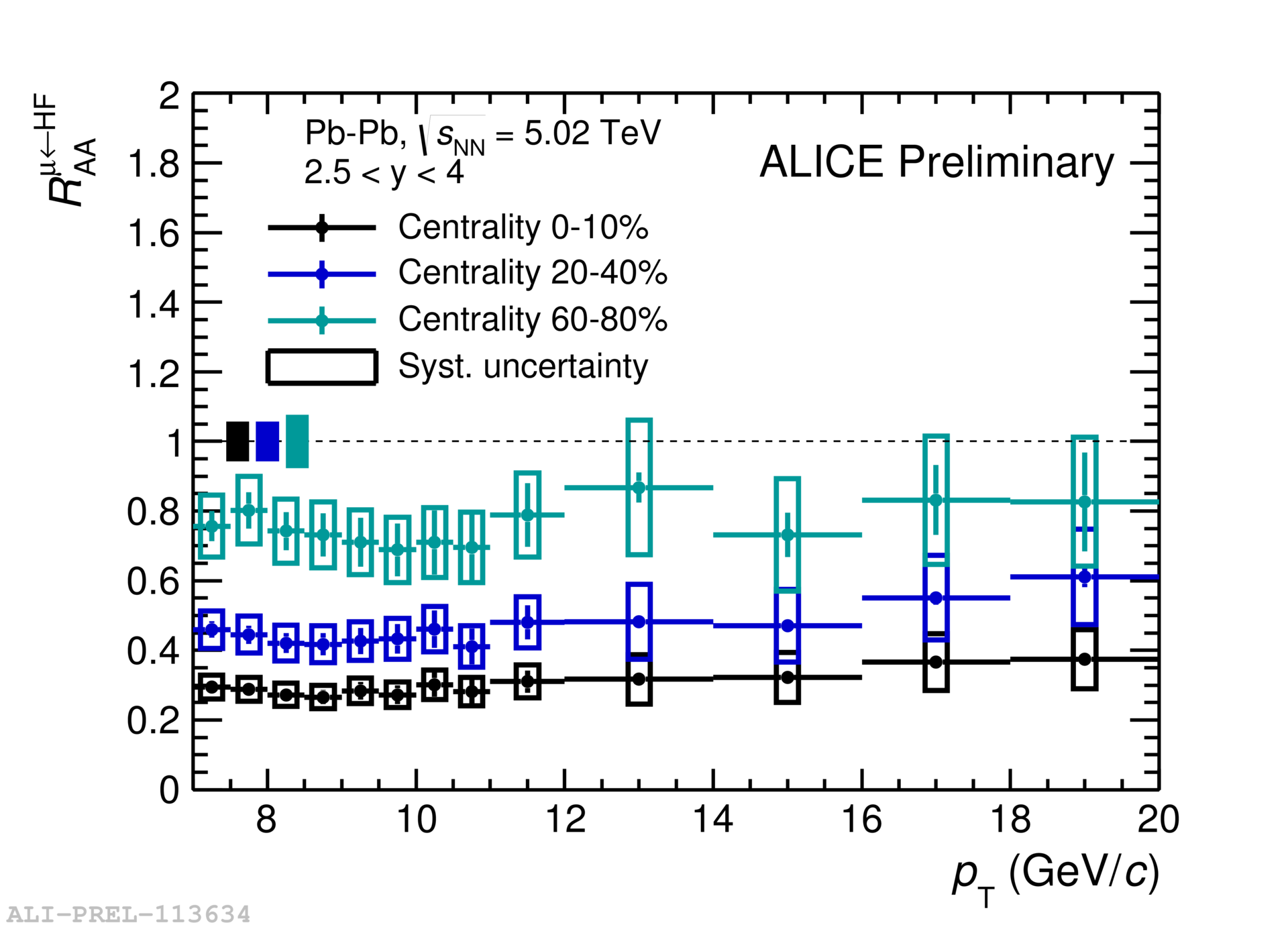}
    \includegraphics[width=8cm]{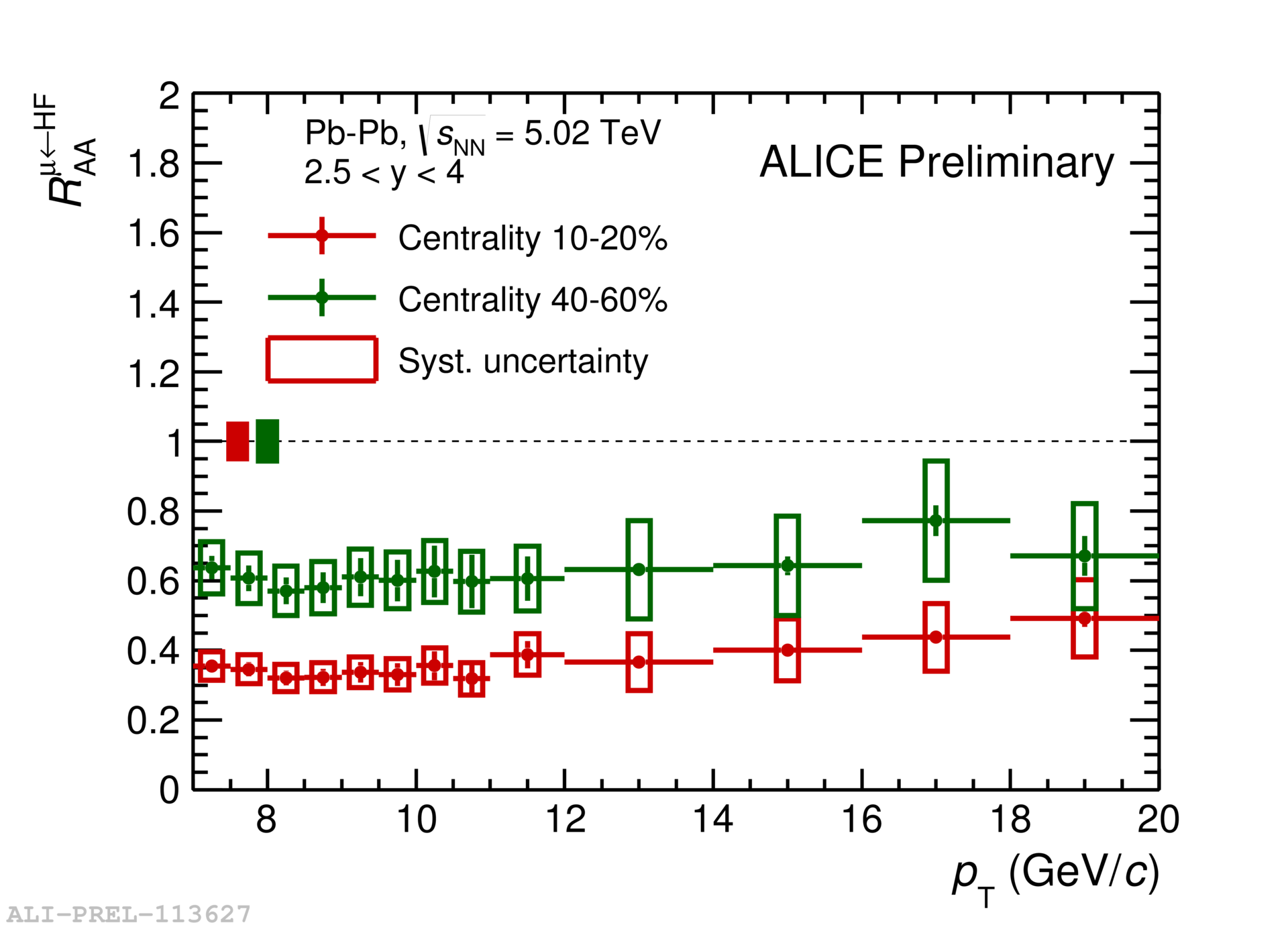}
    \caption{$p_{\rm T}$-differential $R_{\rm AA}$ of heavy-flavour hadron decay muons at forward rapidity in different centrality classes.}
    \label{Fig:RAAmuon}
  \end{figure}

 \begin{figure}
    \centering
    \includegraphics[width=8cm]{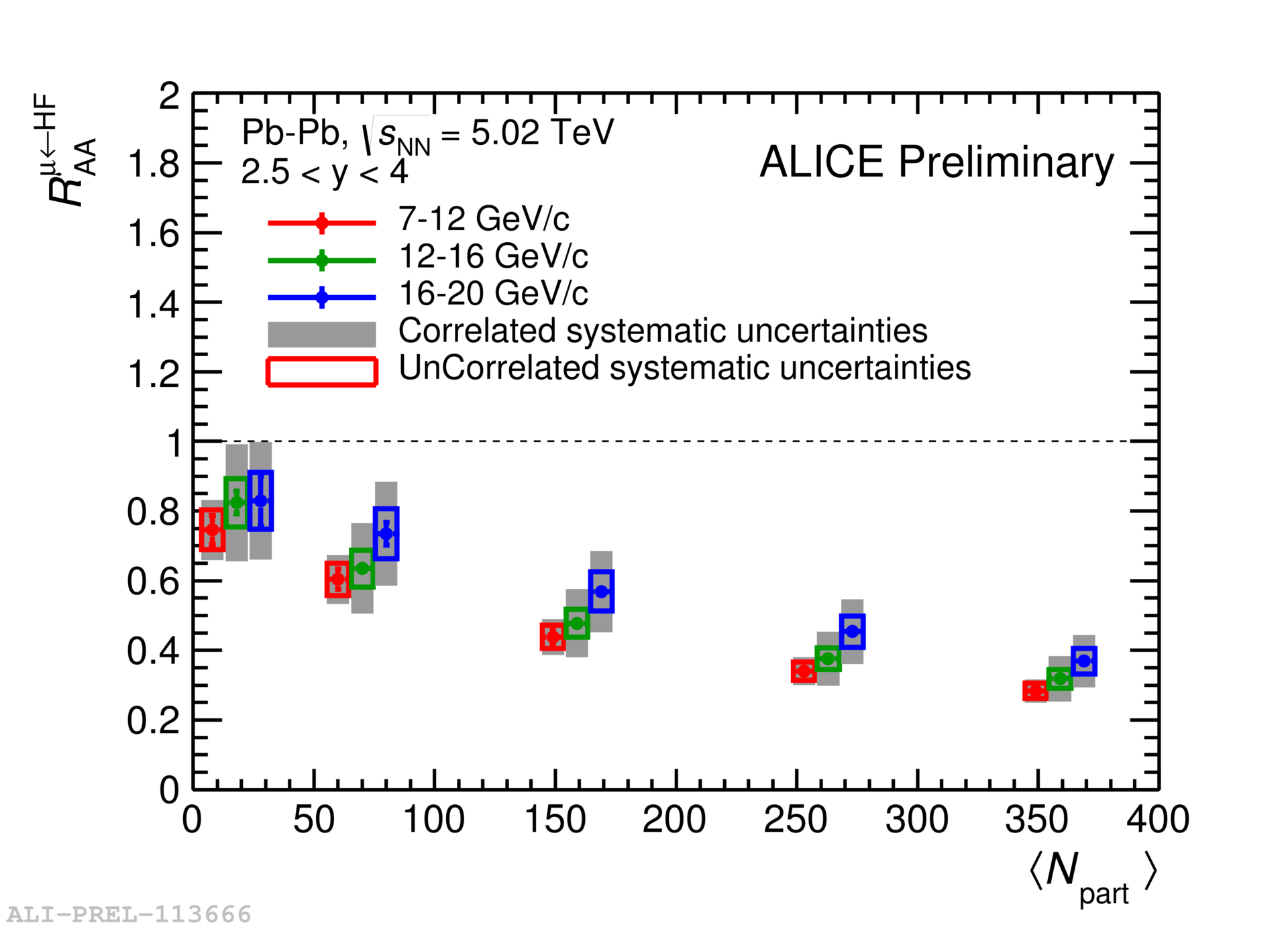}
    \caption{$R_{\rm AA}$ of heavy-flavour decay muons as a function of the mean number of participating nucleons for different transverse momentum ranges.}
    \label{Fig:Integral_RAAmuon}
  \end{figure}

   The background of $\pi$ and K decay muons that needs to be subtracted is evaluated by a data-tuned
   Monte-Carlo cocktail. The fraction of $\pi$ and K decay muons is about 3\% in $7 < p_{\rm T} < 10 ~{\rm GeV}/c$.
   The contribution of W and Z/$\gamma^{*}$ \cite{pub:cms} decay muons in Pb--Pb templates is obtained by
   combining simulations of W and Z/$\gamma^{*}$ decay muons in pp, pn, np and nn collisions with
   POWHEG.
   The fraction of W and Z/$\gamma^{*}$ decay muons increases with $p_{\rm T}$ and is about 38\% (19\%) in the 0--10\% (60--80\%) centrality class at $p_{\rm T}$ = $20 ~{\rm GeV}/c$.

  The study of in-medium effects with the $R_{\rm AA}$ observable requires a pp
  reference. The latter was obtained from the published $p_{\rm T}$-differential cross section of  heavy-flavour hadron decay muons at $\sqrt{s} = 7 ~{\rm TeV}$ measured in $2 < p_{\rm T} < 12 ~{\rm GeV}/c$ ~\cite{pub:pp7} scaled to $\sqrt{s} = 5.02 ~{\rm TeV}$ using simulations based on  Fixed-Order-Next-to-Leading-Log (FONLL) ~\cite{pub:energyscale} calculations. The $p_{\rm T}$-differential cross section of heavy-flavour decay muons in $p_{\rm T} > 12 ~ {\rm GeV}/c$ is obtained by
  scaling FONLL predictions according the ratio between data and FONLL.

  The systematic uncertainty on the nuclear modification factor of muons from heavy-flavour hadron decays
  is affected by the systematic uncertainties of the inclusive muon yield, the background subtraction,
  the pp reference and the normalization procedure. The final
  systematic uncertainty on the $p_{\rm T}$-differential $R_{\rm AA}$ is obtained by adding in quadrature these contributions. It varies from about  11\% ($p_{\rm T}$ = 7 GeV/$c$) to about 22\% ($p_{\rm T}$ = 20 GeV/$c$) in central collisions.

\section{Results}

  Figure~\ref{Fig:RAAmuon} shows the $p_{\rm T}$-differential $R_{\rm AA}$ of heavy-flavour decay muons in the $p_{\rm T}$ interval $7< p_{\rm T} <20 ~{\rm GeV}/c$, in the centrality classes 0--10\%, 10--20\%, 20--40\%, 40--60\% and 60--80\%. The vertical error bars represent the statistical uncertainty. The open boxes include the systematic uncertainty sources except the systematic uncertainty on normalization which is shown as a filled box at $R_{\rm AA} = 1 $.
  A stronger suppression is observed in central collisions compared to peripheral collisions, with no significant $p_{\rm T}$ dependence within uncertainties. The suppression reaches a factor of about three in the 10\% most central collisions.

   \begin{figure}
    \centering
    \includegraphics[width=8cm]{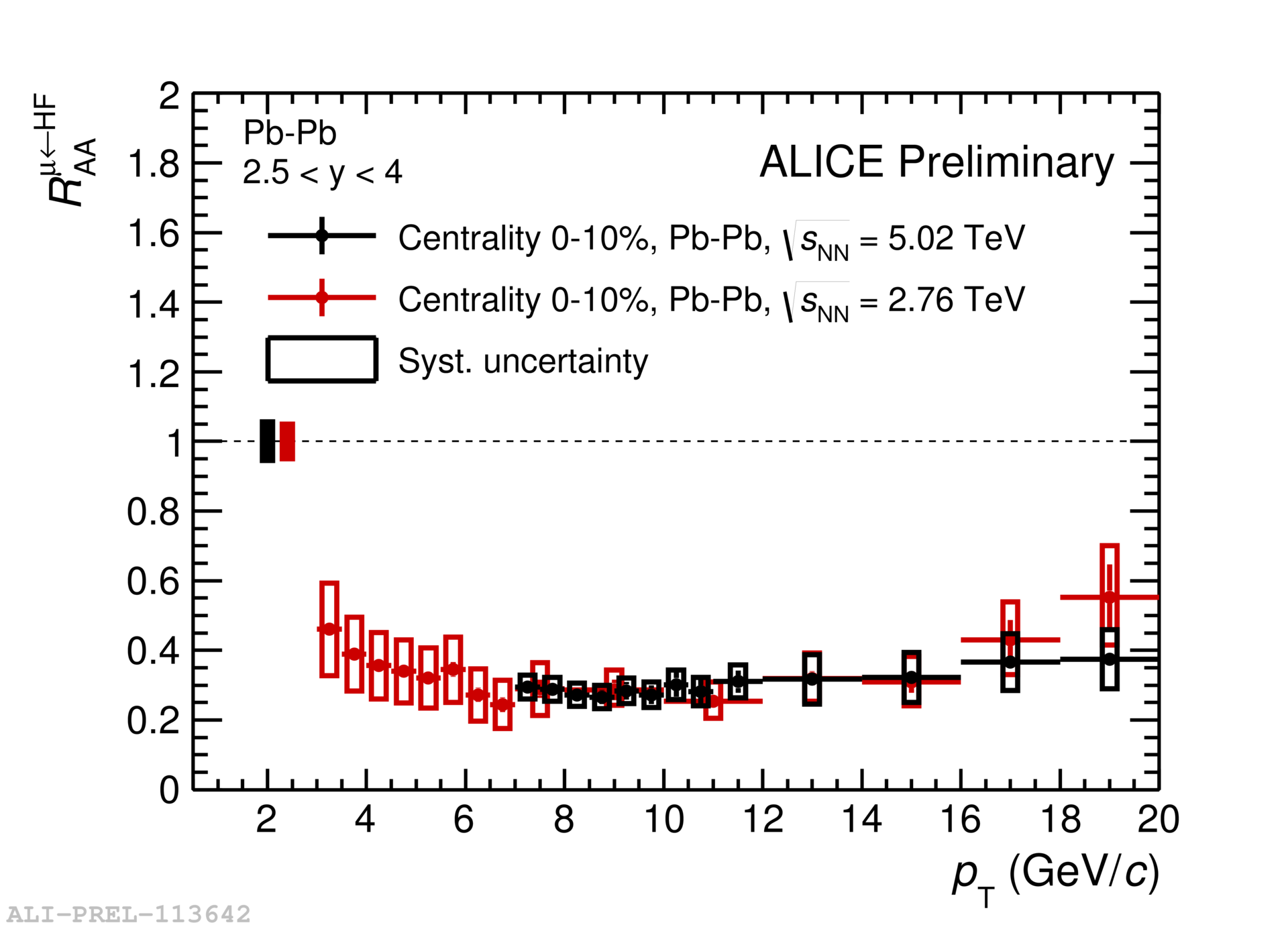}
    \caption{$p_{\rm T}$-differential $R_{\rm AA}$ of heavy-flavour hadron decay muons at forward rapidity in Pb-Pb collisions at $\sqrt{s_{\rm NN}}$ = 5.02 compared with the measurement at $\sqrt{s_{\rm NN}}$ = 2.76 TeV.}
    \label{Fig:RAAmuon_run1_run2}
  \end{figure}

   \begin{figure}
    \centering
    \includegraphics[width=8cm]{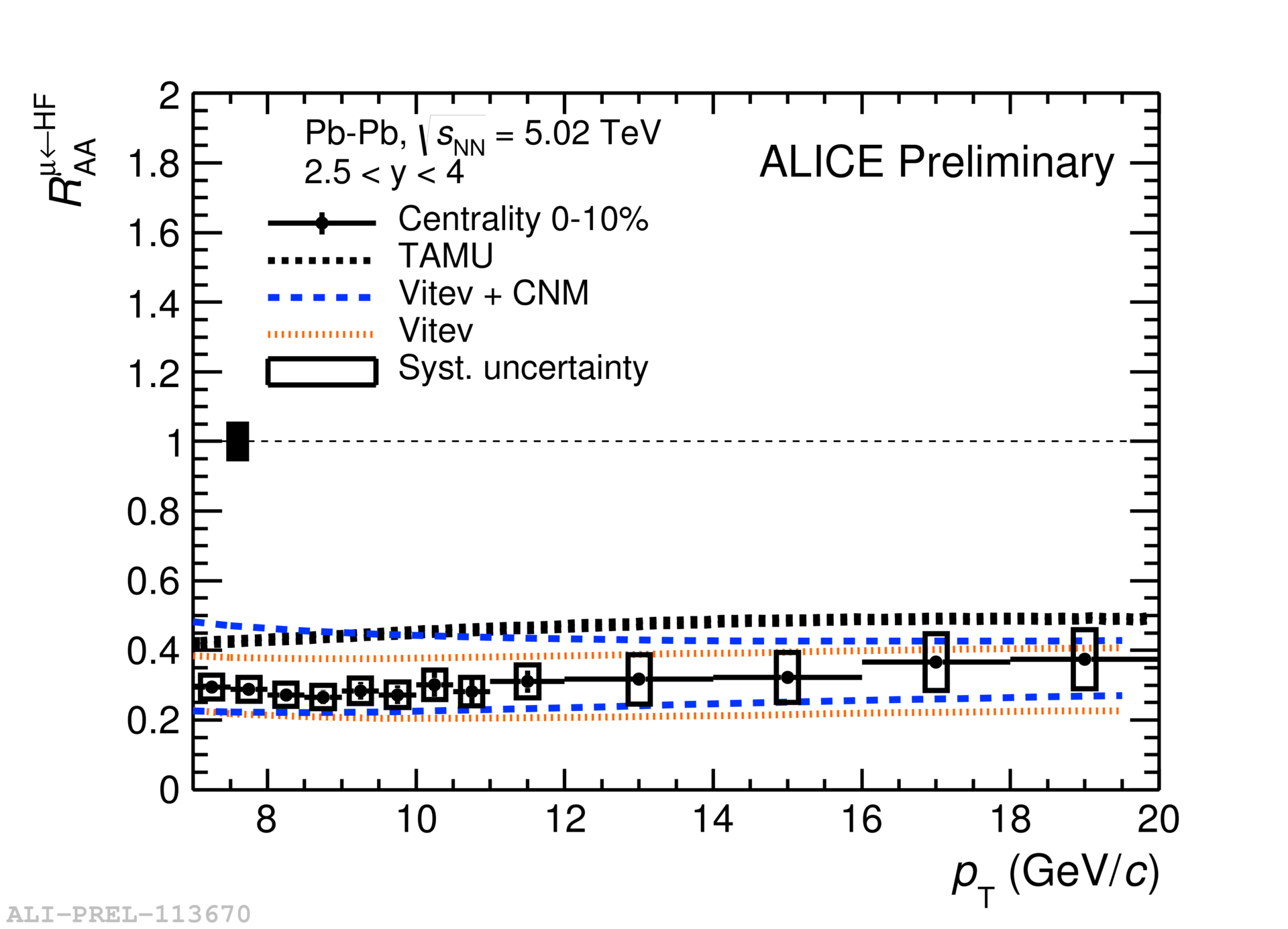}
    \includegraphics[width=8cm]{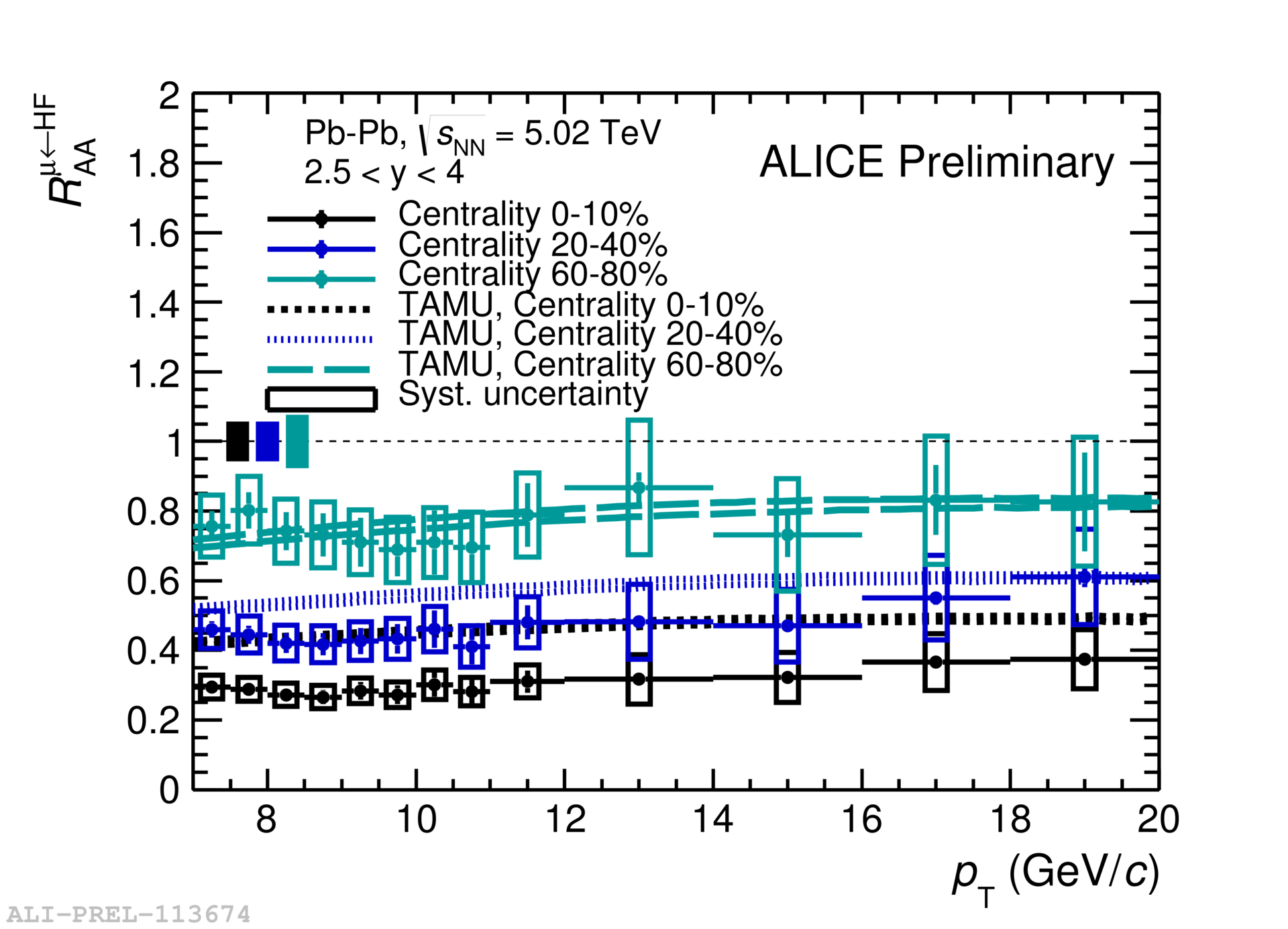}
    \caption{$R_{\rm AA}$ of muons from heavy-flavour hadron decays in $2.5 < y < 4$, compared with different model predictions.}
    \label{Fig:RAAmuon_model}
  \end{figure}

  Figure~\ref{Fig:Integral_RAAmuon} shows the centrality dependence of $R_{\rm AA}$ of heavy-flavour hadron decay muons which is studied as a function of the average number of participating nucleons ($<N_{\rm part}>$).
  The suppression increases with increasing centrality in the domain $7 < p_{\rm T} < 20 ~{\rm GeV}/c$.
  The results are compatible with those obtained in Pb--Pb collisions at $\sqrt{s_{\rm NN}} = 2.76$ TeV~\cite{pub:PbPb2.76}, although in a slightly different $p_{\rm T}$ interval.

  Figure~\ref{Fig:RAAmuon_run1_run2} presents a comparison of the $p_{\rm T}$-differential $R_{\rm AA}$ of heavy-flavour hadron decay muons at forward rapidity in Pb--Pb collisions at 5.02 TeV and 2.76 TeV. A similar suppression is observed at 5.02 TeV and at 2.76 TeV within uncertainties.

  Figure~\ref{Fig:RAAmuon_model} presents a comparison of the $R_{\rm AA}$ of muons from heavy-flavour decays in $2.5 < y < 4$ with transport models in different centrality classes. The Vitev model~\cite{pub:Vitev} (Fig. 4, top) describes the $R_{\rm AA}$ of heavy-flavour hadron decay muons in central collisions. The TAMU model~\cite{pub:TAMU} tends to overestimate the $R_{\rm AA}$ of muons from heavy-flavour hadron decays in central collisions and reproduces the measurement in peripheral collisions.
  These $R_{\rm AA}$ measurements at $\sqrt{s_{\rm NN}} = 5.02~{\rm TeV}$ provide new constraints on energy loss models.

  In addition to final-state effects where in-medium
  energy loss would be dominant, initial-state effects could influence the $R_{\rm AA}$ measurement. In
  the kinematic range relevant for the study of the production of muons from heavy-flavour hadron decays, the main expected effects are
  nuclear modification of parton distribution functions~\cite{pub:RpA_M1}, energy loss~\cite{pub:RpA_M2} and
  $k_{\rm T}$ broadening via multiple soft scatterings in the initial state~\cite{pub:RpA_M3}.
  It is worth noting that since the corresponding measured  $R_{\rm pPb}$ is consistent with unity over the whole $p_{\rm T}$ range~\cite{pub:RpA_data},
  the suppression observed at high $p_{\rm T}$  in central Pb-Pb collisions results from final-state effects related to parton energy loss~\cite{pub:deadcone} in-medium.

\section{Summary}

  We presented the first measurement of the nuclear modification factor of open heavy flavours with ALICE via single muons measured at forward rapidity ($2.5 < y <4$) in a wide $p_{\rm T}$ range ($7 < p_{\rm T} < 20 ~{\rm GeV}/c$) in Pb--Pb collisions at $\sqrt{s_{\rm NN}} = 5.02 ~{\rm TeV}$ collected in the LHC Run2.
  A strong suppression in the 10\% most central collisions reaching a factor of about three in $7 < p_{\rm T} < 12$ ~${\rm GeV}/c$ is observed.
  Results are compatible within uncertainties with those obtained at $\sqrt{s_{\rm NN}} = 2.76~{\rm TeV}$.
  The measured suppression is due to final-state effects.
  The results of $R_{\rm AA}$ in Pb--Pb collisions at $\sqrt{s_{\rm NN}} = 5.02 ~{\rm TeV}$ in various centrality classes may provide important constraints to the models.




\begin{flushleft}
    \bf{Acknowledgement}
\end{flushleft}

This work is supported partly by the NSFC 11375071, the National Basic Research Program of China (2013CB837803) and NSFC IRG11521064 and 11475068.

\nocite{*}
\bibliographystyle{elsarticle-num}
\bibliography{jos}







\end{document}